\documentclass{article}
\usepackage{romp40}
\usepackage{graphicx,color}
\title{QUANTUM FLUCTUATIONS AND GEOMETRY: FROM GRAPH COUNTING TO RICCI FLOW}
\author{ Mauro Carfora \\ Dipartimento di Fisica Nucleare e Teorica, Via Bassi 6, 27100 Pavia, Italy\\
 and\\
 Istituto Nazionale di Fisica Nucleare, Sezione di Pavia, Via Bassi 6, 27100 Pavia, Italy \\ e-mail:mauro.carfora@pv.infn.it
\\[2ex]
         Stefano Romano 
                      \\ SISSA, via Beirut 2-4, 34014 Trieste, Italy \\ e-mail: 	sromano@sissa.it }

\begin{document}

\maketitle
\begin{abstract}
     We discuss in rather general terms quantum field theories dealing with spaces of maps between Riemannian manifolds.  In particular we explore the well--known connection between the renormalization group flow for non--linear sigma models and the Ricci flow. This is an expanded version of the invited talk that one of us (M.C.) has presented at 
The Jubilee 40th Symposium on Mathematical Physics "Geometry and Quanta" (Torun, June 25-28, 2008)
\end{abstract}


\section{Introduction}
 Perhaps mathematical imagination is a part of our technology, and with such a spirit
we shall examine some aspects of the relation between Riemannian geometry and Quantum Field Theories (QFTs). Our  emphasis will be on the connection between QFT and Ricci flow, rewieving, with a geometrical spirit, the well--known QFT computation leading to Ricci flow. We shall confine ourselves to the most elementary facts, not as a device to keep the audience at arm's length but rather out of ignorance of the full depth of a subject which bounces and surfs on geometric analysis, all by itself, and makes it look easy.\\
\\
Notationwise, we shall generically denote by $\Sigma$ and $M$ Riemannian manifolds and deal with spaces of maps $\phi: \Sigma\longrightarrow M$
parametrized by a set of couplings $\mathcal{C}$. Note that $\mathcal{C}$ itself may (and it will) be an infinite dimensional space of geometrical origin. Note also that in such a general setting one must be imaginative enough not to assume a priori any strong regularity assumptions on the maps $\phi: \Sigma\to M$, and yet still pretend that one can give a reasonable mathematical meaning to the spaces $Map\;(\Sigma,M):=\left\{\phi: \Sigma\longrightarrow M\right\}$ and $\mathcal{C}$. By definition $dim\,\Sigma$ is the \emph{dimension} of the given QFT. In a classical field theory, ambiented in  $Map(\Sigma,M)\times\mathcal{C}$,  we usually deal with a given local action functional describing the energetics of the field $\phi: \Sigma \to M$, 
$$
\mathcal{S}: Map(\Sigma,M)\times\mathcal{C}\longrightarrow {R}
$$ 
$$
(\phi,{\alpha})\longmapsto \mathcal{S}[\phi;\alpha]\;,
$$
where $\alpha\in\mathcal{C}$ denote a fixed set of couplings. In QFT it is more appropriate, for reasons that will be discussed momentarily, to consider any such action as a point in a formal space $\mathcal{ACT}\,\,[Map(\Sigma,M)\times\mathcal{C}]$ of functionals on  $Map(\Sigma,M)\times\mathcal{C}$. We may identify $\mathcal{ACT}\,\,[Map(\Sigma,M)\times\mathcal{C}]$ with the space of actions associated with natural Lagrangians on $Map(\Sigma,M)\times\mathcal{C}$, \emph{i.e.}, functions $(\phi,{\alpha})\longrightarrow \mathcal{L}(\phi,{\alpha})$ from  $Map(\Sigma,M)\times\mathcal{C}$ to the space of (smooth) functions $C^{\infty }(\Sigma,{R})$ on $\Sigma $, which are invariant under the diffeomorphism groups $\mathcal{D}iff(\Sigma )$, $\mathcal{D}iff(M)$ and depend on some finite order jet of geometrical fields defined on $\Sigma $ and $M$. One typically assumes that $\mathcal{S}[\phi;\alpha]=\int_{\Sigma }\mathcal{L}(\phi,{\alpha})\,d\mu _{\Sigma }$, where $d\mu _{\Sigma }$ is a measure on $\Sigma $. As a matter of fact, a QFT is naturally associated to the \emph{orbit} of the given classical $\mathcal{S}[\phi;\alpha]$ generated, in $\mathcal{ACT}\,\,[Map(\Sigma,M)\times\mathcal{C}]$, by a (semi)group whose existence is forced upon us in order to give a sensible physical meaning to the quantization procedure. This is a long story that, out of necessity, we collapse in a nutshell. It starts by recalling that  a (Euclidean) QFT is characterized by the correlations among the values $\{\phi(x_{i})\}\in (M)^{k}$ that the fields may attain at distinct marked points $x_{1},\ldots,x_{k}\,\in\Sigma$ , induced by a suitable family of $\mathcal{C}$--dependent (probability) measures on $Map(\Sigma,M)$.
These correlations are provided by functional integrals of the form
\begin{equation}
Z\left[\phi(x_{i});\alpha \right]\doteq \int_{\{Map(\Sigma,M)\}}D_{\alpha}[\phi]\,(\phi(x_{1})\ldots \phi(x_{i})\ldots)\;  e^{-S[\phi; \alpha] }\;,
\label{correlations}
\end{equation}
where $D_{\alpha}[\phi] $ is a (typically non--existant) functional measure on $Map(\Sigma,M)$, possibly depending from the couplings $\alpha\in\mathcal{C}$.\\

The somewhat fanciful expression (\ref{correlations}) is defined in terms of quantities none of which makes sense by itself: as already stressed, there is no natural measure such as $D_{\alpha}[\phi] $ on $Map(\Sigma,M)$, moreover the action $S[\phi; \alpha]$ is typically divergent on the class of maps in $Map(\Sigma,M)$ which,  allegedly, should be typically sampled by $D_{\alpha}[\phi] $. Notheless, if we give 
(\ref{correlations}) some degree of acceptance, then the geometrical picture which emerges is quite non trivial already in the $0$--dimensional case where $\Sigma={p}$, a point, and 
$\phi:\Sigma\longrightarrow M\,,\; p\longmapsto \phi(p)$ so that $ Map\;(\Sigma,M):=\left\{\phi: \Sigma\longrightarrow M\right\}\simeq 
M$, (see \cite{Gaw} and the inspiring analysis by P. Etingof in \cite{Eti}, from which we have drawn the following examples). In such a case, the action functional $\mathcal{S}: Map(\Sigma,M)\times\mathcal{C}\longrightarrow {R}$ becomes a scalar function on $M\times\mathcal{C}$, \,\,$\mathcal{S}: M\times\mathcal{C}\longrightarrow {R}$ and
$$
Z[\alpha,\hbar]=\int_{\{Map(\Sigma,M)\}}D[\phi] \; e^{{-\frac{S[\phi;\alpha]}{\hbar}}}=\hbar^{-\frac{n}{2}}\;\int_{M}\; e^{-\frac{S[\phi; \alpha]}{\hbar}} \,\,d\mu_{M}\;, 
$$
$$
S[\phi;\alpha]=\frac{B(\phi,\phi)}{2}+\sum_{m\geq 0}\alpha_{m}\,\frac{B_{m}(\phi,...,\phi)}{m!}
$$
where we have introduced explicitly $\hbar$, the Planck constant, and developed $S[\phi;\alpha]$ in formal Taylor series in powers of $\alpha$, (\emph{i.e.}, we are considering, rather presciently, $S[\phi;\alpha]$ as a deformation of a free--theory). By steepest descent, (or stationary phase), integrals of this type localize, as the coupling $\alpha$ varies, to the critical  points of the function $S[\phi,\alpha]$.
They are parametrized by decorated graphs, and one can write (see \emph{e.g.}, Theor. 3.3, p.11 of \cite{Eti}
$$
Z[{\color{black}\alpha},\hbar]=\frac{(2\pi)^{\frac{n}{2}}}{\sqrt{\det\,B}}\;\sum_{(n_0,n_1,...)}\left(\prod_{i}\,\alpha_{i}\right)\,\sum_{\Gamma\in G(n_0,n_1,...)}\frac{\hbar^{b(\Gamma)}}{\left|Aut\,(\Gamma)\right|}\;F_{\Gamma}\;,
$$
where $G(n_0,n_1,...)$ denotes the set of isomorphism classes of graphs $\Gamma$, without external vertices, with $n_{k}$\,\,\, $k$--valent vertices, $\left|Aut\,(\Gamma)\right|$ is the order of the automorphisms group of $\Gamma$, $b(\Gamma)\doteq \sharp Edges(\gamma)-\sharp Vertices(\gamma)$, and finally $F_{\Gamma}$ is a suitably defined Feynman amplitude for $\Gamma$. By selecting the target manifold $M$, 0--Dim QFT can be used to solve sophisticated enumerative problems in geometry (again, this is nicely described in \cite{Eti}, \S 4 and \S 5). For instance if $M$ is the space of $N\times N$
hermitian matrices $A$ and 
$$
S[A;\alpha]=\frac{Tr\;A^{2}}{2}-\sum_{m\geq 0}\alpha_{m}\,\frac{Tr\;A^{m}}{m}\;,
$$
then (\cite{Eti}, Theor. 4.5) 
$$
{\color{black}\ln\;Z[A;{\color{black}\alpha}]}=\sum_{(n_0,n_1,...)}\left(\prod_{i}\,\alpha_{i}\right)\,\sum_{\Gamma\in G(n_0,n_1,...)}\frac{N^{2-2g(\Gamma)}\;\hbar^{b(\Gamma)}}{\left|Aut\,(\Gamma)\right|}\;,
$$
where $g(\Gamma)$ is the genus of the ribbon graph,  can be use to count  ribbon graphs parametrization of Riemann surfaces, thus establishing 
a deep connection between $0$--dim QFT and the topology of the moduli space of Riemann surfaces \cite{Harer}. One--dimensional QFT, \emph{i.e.}, 
$\Sigma= {R},{S}^{1}$, is synonymous of quantum mechanics on the line or on the circumference. In this case 
$$
Map(\Sigma,M)=\mathcal{P}(t,x;t',x'):=
\left\{\phi\in C^{\infty }(\Sigma,M)\,\,|\,\,\phi(t)=x,\;\phi(t')=x'  \right\}\;,
$$
and the action takes the form of the standard Lagrangian based action of particle mechanics, $S[\phi(t)]=\int_{\Sigma}\;\mathcal{L}(\phi(t),\dot{\phi}(t))\,dt$.
The resulting QFT lends itself to a well--known functional integral representation \cite{Feynman} \cite{Glimm}, and we are called to evaluate normalized (Euclidean) correlation functions of the form
$$
{\color{black}G_{n}[{\color{black}t_{1},...,t_{n}}]}\doteq \frac{\int_{\mathcal{P}(t,x;t',x')}\,\phi(t_1)\ldots \phi(t_n)\,e^{-S_E[\phi]}\,D[\phi]}{\int_{\mathcal{P}(t,x;t',x')}\,e^{-S_E[\phi]}\,D[\phi]}\;,
$$
where $S_E[\phi]$ denotes the Euclidean action obtained from $S[\phi(t)]$ by the Wick rotation $t\mapsto i t$. Such correlation functions are again
governed by an expansion in $k$--valent ($k\geq 3$) graphs $\Gamma\in G^{*}_{\geq 3}(n)$, (with $n$ external legs), decorated by distributions, $l_{i}\doteq\delta(t-t_{i})$ (external legs), and Green functions acting between function spaces (internal edges). The corresponding Feynman amplitudes $F_{\Gamma}(l_{1},...,l_{n})$ parametrize expectations of jointly Gaussian random variables, (\cite{Eti} \S 7.3),
$$
{\color{black}G_{n}[{\color{black}t_{1},...,t_{n}}]}=\sum_{\Gamma\in G^{*}_{\geq 3}(n)}\frac{\hbar^{b(\Gamma)}}{\left|Aut\,(\Gamma)\right|}
\;{\color{black}F_{\Gamma}}({\color{black}l_{1},...,l_{n}})\;,
$$
describing a Wiener process on $M$. On the geometrical side, one may say that in the case of 1-Dim QFT (Euclidean), the functional integral feels how the fluctuations, as we change the lenght scale $\Delta t$ in $\Sigma$, affects the random paths  $t\mapsto \phi(t)$ in $M$. In such a sense one is not surprised by the fact that 1--Dim QFT can be used to  probe and reconstruct the geometry of $M$: we are actually dealing with a quantization of the geodesic flow on $M$. Indeed,  the triple $(L^{2},\Delta,\mathcal{A})$, where $\mathcal{A}$ is the algebra of operators
in $L^{2}$ acting by multiplication by smooth functions, allows to reconstruct the Riemannian geometry of $M$, (see e.g. \cite{FG} for a nice discussion of such a topic), and we may identify Riemannian Geometry of $M\Longleftrightarrow $ Quantum Mechanics of test particles on $M$. \\

\section{The renormalization group flow}

Until now we have recalled examples where quantum fluctuations, as described by a QFT, probe aspects of the geometry 
of $(\Sigma, M)$ without, however, directly affecting it. Thus, one may wonder if and to what extent quantum fluctuations of the fields $\phi:\Sigma\to M$ can deform the geometry of the pair $(\Sigma,M)$. In order to answer such a question we need to keep under control the behavior of both the maps  $\phi: \Sigma\to M$ and the couplings $\alpha\in\mathcal{C}$ as we vary the scale of distances in $\Sigma$, (the only scale of measurement significant in a relativistic quantum theory). The necessity of such control is already present in $1$--dimensional QFT, but it becomes a fundamental issue for $2$--dim QFT.  As a matter of fact,   
a basic ingredient of any such a QFT is the search for a set of transformations, (\emph{renormalization group flow}),
\begin{eqnarray}
\mathcal{RG}_{t}:\,[Map(\Sigma,M)\times\mathcal{C}]&\longrightarrow& [Map(\Sigma,M)\times\mathcal{C}]\label{reflow}\\
(\phi,\alpha)\;\;\;\;\;\;\;&\longmapsto& \mathcal{RG}_{t}(\phi,\alpha)=({\phi}_{t};{\alpha}(t))\;,\nonumber
\end{eqnarray}
which, as we vary the scale $t$ at which we probe the Riemannian surface $\Sigma$, allow to tame the energetics of the fluctuations of the fields $\phi : \Sigma\to M$, and tune the couplings $\alpha\mapsto\alpha(t)$, accordingly. In order to describe this procedure in general terms, select two scales of distances, say $\Lambda^{-1}$ and $\Lambda'^{-1}$, (one can equivalently interpret $\Lambda$ and $\Lambda'$ as the respective scales of momentum in the spectra of field fluctuations), with   $\Lambda'^{-1}>\Lambda^{-1}$.  The general idea, central in K.G. Wilson's analysis of the the renormalization group flow, is to assume that if $S[\phi; \alpha]\in \mathcal{ACT}\,\,[Map(\Sigma,M)\times\mathcal{C}]$ describes the theory at  a cut--off scale $\Lambda^{-1}$, then  there is a map 
$$
{\widetilde\mathcal{RG}}_{\Lambda\Lambda'}:\mathcal{ACT}\,\,[Map(\Sigma,M)\times\mathcal{C}]\to\mathcal{ACT}\,\,[Map(\Sigma,M)\times\mathcal{C}]\;,
$$
such that the action $S'[\phi'; \alpha']\doteq \left({\widetilde\mathcal{RG}}^*_{\Lambda\Lambda'}\,S\right)[\phi; \alpha]$ provides the effective theory at scale $\Lambda'^{-1}$, obtained upon suitably averaging field--fluctuations in moving from the distance scale $\Lambda^{-1}$ to the scale $\Lambda'^{-1}$. Such a map is required to satisfy the semigroup property   ${\widetilde\mathcal{RG}}_{\Lambda\Lambda''}={\widetilde\mathcal{RG}}_{\Lambda'\Lambda''}\circ {\widetilde\mathcal{RG}}_{\Lambda\Lambda'}$ for all $\Lambda''<\Lambda'$. This formal (semi)-flow, if exists, induces a corresponding flow (field redefinitions) on the measure space $\left(Map(\Sigma,M); D_{\alpha}[\phi]\right)$, and a flow in the space of couplings $\mathcal{C}$. The idea is roughly the following: suppose that, at least for $\left(\frac{\Lambda'}{\Lambda}\right)$ small enough, we can put the functional measure $\widetilde{\mathcal{RG}}_{\Lambda\Lambda'}^{*}\,(D_{\alpha}[\phi])\;  e^{-\left(\widetilde{\mathcal{RG}}^*_{\Lambda\Lambda'}\,S\right)[\phi; \alpha]}$ in the same form as our original measure $D_{\alpha}[\phi]\;  e^{-S[\phi; \alpha] }$, except for a small modification of the couplings $\alpha$. Explicitly, let $\Lambda'=e^{-\epsilon}\,\Lambda$, with $0<\epsilon<1$, and assume that for every such $\epsilon$ there exists a corresponding coupling $\alpha+\delta\,\alpha$ such that the following identity holds
\begin{equation}
\mathcal{RG}_{\epsilon}^{*}\,(D_{\alpha}[\phi])\;  e^{-\left(\mathcal{RG}^*_{\epsilon }\,S\right)[\phi; \alpha]}= D_{\alpha+\delta\alpha}[\phi]\;  e^{-S[\phi; \alpha+\delta\alpha] }\;,
\label{infrenor}
\end{equation}
where we have denoted $\mathcal{RG}_{\epsilon}$ the action of the map $\widetilde{\mathcal{RG}}_{\Lambda\Lambda'}$ for $\Lambda'=e^{-\epsilon}\,\Lambda$.
In other words, we assume that an infinitesimal change in the cutoff can be completely \emph{absorbed} in an infinitesimal change of the couplings. If this equation  is valid at least to some order in $\epsilon$,  we can iteratively use it to see how $\alpha$ is affected by a finite change of the cutoff. If we set $t\doteq\,-\ln\,\left(\frac{\Lambda'}{\Lambda}\right)$, then the map $\mathcal{RG}_{t}$ so induced by $\widetilde{\mathcal{RG}}_{\Lambda\Lambda'}$ on $[Map(\Sigma,M)\times\mathcal{C}]$, as $t$ varies, is the renormalization group flow $\mathcal{RG}_{t}$ introduced above, (see (\ref{reflow})).
The basic issue concerns in which sense, under the action of $\mathcal{RG}_{t}$,  the functional $S[\phi;\alpha]$ defines a cut-off  independent QFT, \emph{i.e.}, if we can take the $\Lambda\to\infty$ limit. An admittedly formal, yet geometrically elegant, answer is that a QFT is characterized by the given 
action  $\mathcal{S}$ only if the associated functional measure $D_{\alpha}[\phi]\;  e^{-S[\phi; \alpha] }$
has natural transformation properties under $\mathcal{RG}_{t}$, \emph{i.e.}, if
\begin{equation}
\int_{\mathcal{RG}_{t}\{Map(\Sigma,M)\}}D_{\alpha(t)}[\phi_t]\;  e^{-S[\phi_t; \alpha(t)] }
= \int_{\{Map(\Sigma,M)\}}\,\mathcal{RG}_{t}^{*}\,(D_{\alpha}[\phi])\;  e^{-\left(\mathcal{RG}_{t}^{*}\,S\right)[\phi; \alpha] }\;,
\label{renorm}
\end{equation}
\emph{holds in the  $t\to -\infty$ limit}. Recall that $\mathcal{RG}_{t}$ is, despite its name, only a semiflow: as the \emph{time $t$} increases we are describing the spectrum of fluctuations of the maps in $Map(\Sigma,M)$ at larger and larger distance scales,   averaging out irrelevant degrees of freedom.  Thus the validity of (\ref{renorm}) in the  $t\to -\infty$ limit is a highly non trivial requirement,  since it is extremely difficult, if not impossible, to go backward in  \emph{time $t$}, by  reversing such an averaging process. This explains why QFTs are conceptually difficult to construct. Whenever this is possible, (\ref{renorm}) says that there correspondingly exists a limit space $\left[\lim_{t\to -\infty}\,\mathcal{RG}_{t}\{Map(\Sigma,M)\}\right]$ of geometrical objects describing the given QFT. Such objects typically do not belong to the original space $Map(\Sigma,M)$, because the backward $\mathcal{RG}_{t}$--flow can be highly singular. To discuss the properties of such a flow, one typically assumes that (\ref{renorm}) holds for any finite scale interval $-\epsilon \leq t \leq \epsilon $, for $\epsilon>0$, and exploit (\ref{infrenor}) by evaluating, along the $\mathcal{RG}_{t}$ map, the flow derivative $\frac{d}{d t}\,Z[\alpha(t)]$ at the generic scale $t$, where 
$$
Z[\alpha(t)]\doteq \int_{\mathcal{RG}_{t}\{Map(\Sigma,M)\}}D_{\alpha(t)}[\phi_t]\;  e^{-S[\phi_t; \alpha(t)] }\;. 
$$
Denoting, from notational ease, $\mathcal{RG}_{t}(\Sigma,M):=\mathcal{RG}_{t}\{Map(\Sigma,M)\}$ we compute
\begin{eqnarray}
\label{zdiff}
&&\frac{d}{d t}\,Z[\alpha(t)]=\lim_{\epsilon \rightarrow 0}\,\frac{1}{\epsilon }\left[\int_{\mathcal{RG}_{t+\epsilon }(\Sigma,M)}D_{\alpha(t)}[\phi_t]\;  e^{-S[\phi_t; \alpha(t)] }\right.\\
&&\left.-\,\int_{\mathcal{RG}_{t}(\Sigma,M)}D_{\alpha(t)}[\phi_t]\;  e^{-S[\phi_t; \alpha(t)] }\right]\nonumber\\
\nonumber\\
&&=\lim_{\epsilon \rightarrow 0}\,\frac{1}{\epsilon }\left[\int_{\mathcal{RG}_{\epsilon }(\mathcal{RG}_{t}(\Sigma,M))}D_{\alpha(t)}[\phi_t]\;  e^{-S[\phi_t; \alpha(t)] }\right.\nonumber\\
&&\left.-\,\int_{\mathcal{RG}_{t}(\Sigma,M)}D_{\alpha(t)}[\phi_t]\;  e^{-S[\phi_t; \alpha(t)] }\right]\nonumber\\
&&=\int_{\mathcal{RG}_{t}(\Sigma,M)}\lim_{\epsilon \rightarrow 0}\,\frac{1}{\epsilon }\left(\mathcal{RG}_{\epsilon }[D_{\alpha(t)}[\phi_t]]\;  
e^{-(\mathcal{RG}_{\epsilon }^*S)[\phi_t; \alpha(t)]}-D_{\alpha(t)}[\phi_t]\;  e^{-S[\phi_t; \alpha(t)] }     \right)
\nonumber\\
&&=-\,\int_{\mathcal{RG}_{t}\{Map(\Sigma,M)\}}\,\beta(\alpha(t))\frac{\partial}{\partial \alpha(t)}\,
\left(D_{\alpha(t)}[\phi_t]\;  e^{-S[\phi_t; \alpha(t)] }\right)\nonumber\;,
\end{eqnarray}
where we have set
\begin{equation}
\beta(\alpha(t)) \doteq -\frac{\partial}{\partial t}\,\alpha(t)\;,
\label{betaflow}
\end{equation}
(the use of the minus sign and of the partial derivative here is somewhat hydiosyncratic), and where we have exploited the semigroup property of the flow and the scaling hypothesis (\ref{infrenor}).
Since the integration is over $\mathcal{RG}_{t}\{Map(\Sigma,M)\}$, we can formally extract the operator $\beta(\alpha(t))\frac{\partial}{\partial \alpha (t)}$   from under the functional integral, and 
 rewrite the relation  (\ref{zdiff}) more synthetically  as 
\begin{equation}
\left\{\frac{d}{d t} +\beta(\alpha(t))\frac{\partial}{\partial \alpha(t)}\right\}\,
Z_t[\alpha]=0\;.
\label{renormflow}
\end{equation}
Note that  the function $\beta(\alpha(t))$ defined by (\ref{betaflow})
can be considered as a \emph{vector field} on the space of couplings $\mathcal{C}$. Roughly speaking (\ref{renormflow}) says is that if we rescale  distances in $\Sigma$ by a factor $e^{t}$ and at the same time we flow in the space of couplings along $\beta$ for a \emph{time} $t$, the theory we obtain looks the same as before.\\
If the theory is, along the lines sketched above, renormalizable by a renormalization of the couplings, many of its properties can be desumed by the analysis of (\ref{betaflow}). This bring us directly to the main player of our talk.

\section{Non--linear $\sigma$--models and Ricci flow}

We now specialize to the nonlinear sigma model. This is a $2$--dim QFT where $\Sigma$ is 2-dimensional Riemannian surface with metric $\gamma=\gamma_{\mu\nu}\,dx^{\mu}\otimes dx^{\nu}$, and the target $M$ is a Riemannian manifold with metric $g_{ik}\,d\phi^{i}\otimes d\phi^{k}$. In particular, we will assume that $\Sigma$ is the flat torus ${T}^2={R}^2/{Z}^2$,  with the metric $\gamma_{\mu\nu}=\delta_{\mu\nu}$. The classical action  associated with a map $\phi:\Sigma\to M$, (assuming that the fields are differentiable at least in the distributional sense, and that  the $L^2(\Sigma)$-norm of the differential $d\phi$ is finite), is defined by
\begin{equation}
S(\phi;a^{-1}\,g) \doteq a^{-1}\,|d\phi|_{L^2(\Sigma)}^2=a^{-1}\,\int_\Sigma\gamma^{\mu\nu}\partial_\mu\phi^i\partial_\nu\phi^j\,g_{ij}\,d\mu_\gamma\;,
\label{sigmamod}
\end{equation}
where $d\mu_\gamma$ is the Riemannian volume element on $(\Sigma, \gamma)$, and where  $a>0$ is a parameter with the dimension of a length squared.  We wrote the action as $S(\phi;a^{-1}\,g)$ in order to  emphasize the fact that at each point $x\in\Sigma$ the metric $a^{-1}\,g(\phi(x))$ plays the role of the coupling constants for the fields $\phi(x)$ of the theory. This suggests that in this case the space of couplings is the infinite-dimensional cone $\mathcal{M}et(M)$ of Riemannian metrics over $M$. However, since the action (\ref{sigmamod}) is invariant under the action of the group of diffeomorphisms of $M$, $\mathcal{D}iff(M)$, we actually have that $\mathcal{C}=\frac{\mathcal{M}et(M)}{\mathcal{D}iff(M)\times R^{+}}$, where $R^{+}$ denotes the group of rescalings defined by $a\mapsto\lambda a$, for $\lambda$ a positive number. Thus $\mathcal{C}$ is the space of Riemannian structures on $M$, modulo overall length rescalings.  
Note that the true dimensionless coupling constant of the theory is
the ratio of the length scale of the target space 
metric $g_{ab}$ (\emph{i.e.}, its squared radius of curvature ${r^{2}_{curv}}$) to $a$. In particular, one may consider a point--like limit,
where the size of the surface $(\Sigma,\gamma)$ is much
smaller than the physical length scale of $(M,g_{ab})$. It is the standard weak--coupling perturbation theory of
the model. If the characteristic scales of $(M,g_{ab})$, (and 
other background fields, see below), are much larger than $a$, then at the 
surface $(\Sigma,h)$ scale the fields are almost flat. To appreciate the meaning of these remarks, recall that  
the action (\ref{sigmamod}) besides being invariant under $\mathcal{D}iff(M)$, it is also invariant under conformal transformations of $(\Sigma,\gamma)$. Its critical point are harmonic maps. In particular, the minima are the constant maps. This implies that when curvature of target Riemannian manifold $(M,g)$ is small as seen by $\Sigma$, (\emph{i.e.}, as in the pointlike limit defined above), 
 (and $\gamma$ flat), the measure  $D_{g}[\phi]\,e^{-S[\phi;a^{-1}\,g]}$ is concentrated
around the constant maps $x\to\phi(x)=\phi_{0}$, and we can
 control the nearly Gaussian fluctuations $\delta\phi$. With a slight abuse of language, it is typical to talk in this case of  perturbation
theory for small $a$, and say that the theory is perturbatively renormalizable in terms
 of the scale parameter $a$. \\
 \\
 A further comment is appropriate here. There are several other terms that one could add to the action (\ref{sigmamod}) while preserving the $Diff(M)$ invariance, (but typically breaking conformal invariance). Some commonly used terms are the \emph{tachion, dilaton} and \emph{topological} terms, defined respectively as
$S_{tach}(\phi)= \int_\Sigma U(\phi)\;d\mu_\gamma$, $S_{dil}(\phi)=\int_\Sigma W(\phi)\mathcal{K}\; d\mu_\gamma$, and
$S_{top}(\phi)=\int_\Sigma \phi^*\omega $, where $U:M\to R$ and $W:M\to R$ are smooth functions on $M$, $\mathcal{K}$ is the Gaussian curvature of $(\Sigma,\gamma)$ and $\omega:M\to \Lambda^{2}(M)$ is a 2-form on $M$. However, for the sake of exposition we will confine our analysis only to the action $S[\phi;a^{-1}\,g]$.
 The perturbative quantization of the corresponding non-linear sigma model is a well--known subject with many connections in geometry. In particular, the landmark papers by Daniel Friedan \cite{DanThesis,DanPRL, DanAnnPhys} on the subject came, at the time, with so many prescient comments on the geometry of the associated $\beta$--flow that they have been, and still are, a rich source of suggestions  for geometric flow theory. For recent results in this connection, the interested reader may consult the fine papers  \cite{Bakas, Bakas2, Bakas3}, \cite{Gegenberg}, \cite{Guenther}, \cite{Oliynyk}, \cite{Tseytlin}. 
\\
\\
It must be noted that the space of maps $Map(\Sigma, M)$ is typically a non--linear functional space, and it is difficult  to implement a renormalization group procedure in such a setting. However, in the weak--coupling limit, where the size of the surface $(\Sigma,\gamma)$ is much
smaller than the physical length scale of $(M,g_{ab})$, (see the remarks above), only fields fluctuating around a constant value $\phi_0\in M$ play a role. It follows that one can work in a geodesic ball $B(\phi_0,r)\subset M$, centered at the point $\phi_0$, with radius $r<\min\left\{\frac{1}{3}\,inj\;(\phi_0),\,\frac{\pi }{6\,\sqrt{K}}  \right\}$, where $K$ is an upper bound to the sectional curvature of $(M,g)$, (we are adopting the standard convention of defining $\pi/2\,\sqrt{K}\doteq\infty$ when $K\leq0$), and $inj\;(\phi_0)$ denotes the injectivity radius of $(M,g)$ at $\phi_0$, (note that $(M,g)$ can be a complete, not necessarily closed, Riemannian manifolds). Under these hypotheses, given $N$ independent copies $\{\phi_k:\Sigma\to B(\phi_0,r)\}_{k=1,\dots,N}$, of the field $\phi:\Sigma\to B(\phi_0,r)$,  one can define their \emph{center of mass} 
\begin{equation}
\psi\doteq cm\, \left\{\phi_1,\ldots,\phi_N \right\}\;, 
\label{cm}
\end{equation}
as the minimizer of the function $F:M\to R$, defined by $F(y)\doteq \frac{1}{2}\,\sum_{k=1}^{N}\,d_{g}^{2}(y,\phi_k)$, where $d_{g}^{2}(\circ ,\circ )$ denotes the distance function in $(M,g)$. Note that if $inj\;(y)>3r$ for all $y\in B(\phi_0,r)$, then the minimizer is unique and $cm\, \left\{\phi_1,\ldots,\phi_N \right\}\,\in B(\phi_0,2r)$\,\, \cite{Chow}. 
 The idea is to describe the QFT, corresponding non--linear sigma model action (\ref{sigmamod}), by extracting the behavior of the (quantum) fluctuations of the maps $\phi:\Sigma\to B(\phi_0,r)$ around the \emph{background (or average) field} $\psi$ defined by the distribution of the center of mass  of a large ($N\to\infty$) number of independent copies of $\phi$ itself. To make this story tellable we shall partly follow the fine presentation of this matter by K. Gawedzki \cite{Gaw}, who does not manufacture cheap answer to such issues.\\
\\
The first key observation is  that if the $N$ fields $\{\phi_k:\Sigma\to B(\phi_0,r)\}$ are distributed on $Map(\Sigma, B(\phi_0,r))\subset Map(\Sigma, B(\phi_0,2r))$ according to the measure  $D_{g}[\phi])\; e^{-(S[(\phi; a^{-1}\,g)] }$, then the associated \emph{background field} $\psi: \Sigma\to B(\phi_0,2r)$, (\ref{cm}), is distributed according to
\begin{equation}
P_N(\psi)= \int_{Map(\Sigma, B(\phi_0,2r))}\,\delta(cm\{\phi_j(x)\}-
\psi(x))\prod_{j=1}^N\,D_{g}[\phi_j]\; e^{-S[(\phi_j; a^{-1}\,g)]}\;,
\label{distribution}
\end{equation}
where $\delta(cm\{\phi_j(x)\}-\psi(x))$ is a formal Dirac measure on $Map(\Sigma, B(\phi_0,2r))$. If one considers the  exponential map, $\exp_{\psi}: T_{\psi}M\to B(\phi_0,2r)$, based at the center of mass $\psi\in B(\phi_0,2r)$, and writes $\phi_k=\exp_{\psi}\,(X_k)$,  $\{X_k\}_{k=1,\ldots,N}\in T_{\psi}M$, then (\ref{distribution}) can be reformulated in terms of fields taking values in the pull--back bundle  $\left.\psi^{*}\,TM\right|_{B(\phi_0,2r)}$. To this end, one introduces $N$ sections $\xi _k:\Sigma\to \psi^{*}\,TM$, with $\;\sum_{j=1}^N\xi_j=0$ (because $\exp_{\psi}\,(\cdot )$ is based at the center of mass $\psi$), and  such that that $X_k=\psi_{*}\,\xi _k$. Thus
\begin{equation}
\phi_k(x) = \exp_{\psi(x)}(\psi_{*}\,\xi_k(x))\;,
\end{equation}
and one can conveniently express (\ref{distribution}) as a functional integral, over the linear space of maps $Map\left(\Sigma, \psi^{*}\,TM\right)\doteq Map\left(\Sigma, \psi^{*}\,TM|_{B(\phi_0,2r)}\right)$, according to
\begin{equation}
\int_{Map\left(\Sigma, \psi^{*}\,TM\right)}\,\delta[\sum_{j=1}^N\xi_j]\prod_{k=1}^N\,e^{-S_{\psi}[\xi_k;\,a^{-1}\,g]}
D^{\psi}_g\,[\xi_k]\;,
\end{equation}
where $S_{\psi}[\xi_k;\,a^{-1}\,g]\doteq S[\exp_{\psi}(\psi_{*}\,\xi_k);\,a^{-1}\,g]$ and $D^{\psi}_g\,[\xi_k]\doteq D_g\,[\exp_{\psi}(\psi_{*}\,\xi_k)]$.
By exploiting the formal Fourier representation of the functional Dirac--$\delta[\circ ]$ one can write
$$
\delta[\sum_{j=1}^N\xi_j]=\int_{Map^{*}(\Sigma, \psi^{*}\,TM)} \left[DJ\right]\,\exp\,i\, \langle J\cdot\sum_{j=1}^N\xi_j \rangle \;,
$$
where the pairing $\langle\circ ,\circ \rangle$ between $Map\left(\Sigma, \psi^{*}\,TM\right)$ and its dual $Map^{*}\left(\Sigma, \psi^{*}\,TM\right)$ is defined by the   $L^{2}$ inner product
$$
\langle J\cdot\sum_{j=1}^N\xi_j \rangle\doteq \int_{\psi^{*}\,TM|_{B(\phi_0,2r)}}\,(\psi^{*}g)_{\mu\nu}\,J^{\mu}\sum_{j=1}^N\xi^{\nu}_j\,\psi^{*}d\mu_{g}\;.
$$
Thus one can eventually express (\ref{distribution}) as
\begin{eqnarray}
\label{Jpartition}
&P_N(\psi)=\int\left[DJ\right]\,\int\,e^{i\,\langle J\cdot\sum_{j=1}^N\xi_j\rangle}\,\prod_{k=1}^N\,e^{-S_{\psi}[\xi_k;\,a^{-1}\,g]}\,D^{\psi}_g\,[\xi_k]\\
\nonumber\\
&=\int\left[DJ\right]\,e^{N\,W_{\psi}(J)}\nonumber\;,
\end{eqnarray}
where we have introduced the characteristic functional of the measure $e^{-S_{\psi}[\eta ;\,a^{-1}\,g]}\,D^{\psi}_g\,[\eta ]$,  $\eta\in\,Map\left(\Sigma, \psi^{*}\,TM\right)$, according to
\begin{equation}
e^{W_{\psi}(J)}\doteq\int_{Map(\Sigma, \psi^{*}\,TM)}\,e^{i\,\langle J\cdot\eta\rangle}\,e^{-S_{\psi}[\eta;\,a^{-1}\,g]}\,D^{\psi}_g\,[\eta]\;.
\end{equation}
Note that one may provide an asymptotic expansion for $W_{\psi}(J)$ by Taylor expanding  $S_{\psi}[\eta;\,a^{-1}\,g]$ around its minimum, (at $\eta=0$), and by separating the  Gaussian measure 
$D^{\psi}_g[\Xi]\doteq{e^{-\frac{1}{2}S_{\mu\nu}^{\psi}\eta^{\mu}\eta^{\nu}}D^{\psi}_g\,[\eta]}\backslash [{\int\,
e^{-\frac{1}{2}S_{\mu\nu}^{\psi}\eta^{\mu}\eta^{\nu}}D^{\psi}_g\,[\eta]}]^{-1}$, 
where $S_{\mu\nu\ldots}^{\psi}$ denotes the covariant derivatives of the action $S_{\psi}[\eta;\,a^{-1}\,g]$ evaluated for $\eta=0$.
In this way we get
\begin{eqnarray}
&&e^{W_{\psi}(J)}=e^{-S_{\psi}[0;\,a^{-1}\,g]}\,\left[\int\,e^{-\frac{1}{2}S_{\mu\nu}^{\psi}[0;\,a^{-1}\,g]\eta^{\mu}\eta^{\nu}}D^{\psi}_g\,[\eta]\right]\;\times \\
\nonumber\\
&&\times \int\,D^{\psi}_g[\Xi]\,e^{i\,\langle J\cdot\eta\rangle}\,e^{-S_{\alpha\mu\nu}^{\psi}[0;\,a^{-1}\,g]\eta^{\alpha}\eta^{\mu}\eta^{\nu}\,-\ldots\,}
\nonumber\;,
\end{eqnarray}
 The expansion in power series of all the exponentials and the resulting term--by--term Gaussian integration provide the formal expression
\begin{equation}
W_{\psi}(J)=-S_{\psi}[0;\,a^{-1}\,g]+\sum_{\Upsilon \in G}\,\frac{(a)^{l(\Upsilon )}}{|Aut(\Upsilon )|}\,F_{\Upsilon }\left(S_{\psi},\,J\right) +\,\ln\,\left(\int\,e^{-\frac{1}{2}S_{\mu\nu}^{\psi}[0;\,a^{-1}\,g]\eta^{\mu}\eta^{\nu}}D^{\psi}_g\,[\eta] \right)\;,
\label{Wfeynman}
\end{equation}
where $G$ denotes the set of isomorphism classes of connected graphs $\Upsilon $ without external lines and with $l(\Upsilon )$ loops.  $F_{\Upsilon }\left(S_{\psi},\,J\right)$ is the Feynman amplitude  of each given $\Upsilon \in G$, computed by associating to $1$--leg vertices the current $J$, to each $n$--leg vertices, $n\geq3$, the interaction $S_{\alpha_1\ldots\alpha_n}^{\psi}[0;\,a^{-1}\,g]$, and to any internal edge the propagator defined by $S_{\mu\nu}^{\psi}[0;\,a^{-1}\,g]$. \\
\\
All these manipulations are, to say the least, formal, but, as customary in QFT, one makes them algorithmically operative by taking the above diagrammatic expansion as the definition of $W_{\psi}(J)$, provided we are able to renormalize the theory to render it finite. This is notoriously a difficult task. However, we are interested in the large $N$ behavior of $P_N(\psi)$ rather than in $W_{\psi}(J)$ itself. According to (\ref{Jpartition}) the large $N$ asymptotics of the distribution $P_N(\psi)$ of the background field $\psi$ is  provided by
$$
P_N(\psi)=e^{N\;\inf_{J}\;W_{\psi}(J)\,+\,o(N)}\;,
$$
where the $\inf$ is over all ${J\in\, Map^{*}(\Sigma, \psi^{*}\,TM)}$. The structure of this asymptotics suggests a second  key observation.  As emphasized by K. Gawedzki in his remarkable lecture notes \cite{Gaw}, $\sup_{J}\;[\langle\zeta,J\rangle- W_{\psi}(J)]$ is the large deviation functional governing the $\mathcal{O}(N)$--fluctuations around $\zeta $, (in our case $\zeta =0$), in the distribution of $\{\xi_j\}_{j=1,\ldots,N}$, as compared to the $\mathcal{O}(\sqrt{N})$ Gaussian fluctuations sampled by the central limit theorem. Since $\sup_{J}\;[\langle\zeta,J\rangle- W_{\psi}(J)]$ is the Legendre transform of $W_{\psi}(J)$, it follows, from standard QFT, that $\sup_{J}\;[\langle\zeta,J\rangle- W_{\psi}(J)]$ can be identified with the \emph{effective action} associated with $W_{\psi}(J)$, \emph{i.e.} with the action functional whose corresponding partition function gives, at tree (classical) level, the full  characteristic functional $W_{\psi}(J)$. According to these remarks it follows that, for the non--linear sigma model (\ref{sigmamod}), the role of a \emph{background field effective action} is played by
$$\Gamma(\psi)\doteq \sup_{J}\;[-\,W_{\psi}(J)]\;.$$
Geometrically, this is the large deviation functional controlling the non--Gaussian fluctuations of the fields $\{\phi_j\}_{j=1,\ldots,N}$,  around a \emph{background (or classical) field} $\psi$ obtained as average of a large number of copies $\{\phi _{j}\}_{j=1,\ldots,N\to\infty}$ of the quantum field itself.\\
\\
The full effective action $\Gamma(\psi)$ can be perturbatively defined, starting from the expansion (\ref{Wfeynman}) of $W_{\psi}(J)$,  by rewriting it in terms of 1--particle irreducible (1PI) graphs, (\emph{i.e.}, in terms of connected graphs  without bridges, where an edge $e$ of a connected graph $\Upsilon $ is said to be a bridge if the graph $\Upsilon \setminus e$ is disconnected). Such a rewriting exploits the well--known result that any connected graph $\Upsilon $ can be uniquely represented as a tree, whose vertices are 1PI irreducible subgraphs, and whose edges are the bridges of $\Upsilon $. From the remarks above, it follows that we need the effective action at tree--level. This can be immediately obtained from the formal expansion (\ref{Wfeynman})   according to 
\begin{eqnarray}
\Gamma_{(0)}(\psi)=S_{\psi}-\sum_{\Upsilon \in G_{1PI}}\,
\frac{(a)^{l(\Upsilon )}}{|Aut(\Upsilon )|}\,F_{\Upsilon }\left(S_{\psi}\right) -a\,\ln\,\left(\int\,e^{-\frac{1}{2}S_{\mu\nu}^{\psi}\,\eta^{\mu}\eta^{\nu}}D^{\psi}_g\,[\eta] \right)\;, 
\label{effaction}
\end{eqnarray}
where $G_{1PI}$ is the set of isomorphism classes of 1PI graphs without $J$--vertices, and $|Aut(\Upsilon )|$ denotes the size of the corresponding automorphisms group.\\
\\
To proceed further, we need the expression of $S_{\psi}[\eta;\,a^{-1}\,g]\doteq S[\exp_{\psi}(\psi_{*}\,\eta);\,a^{-1}\,g]$. Since we are working in a sufficiently small geodesic ball $B(\phi_{0}, 2r)\subset M$, we can safely assume that  $\psi_{*}\eta$ is small and expand the action  keeping terms only up to second order. To this end, consider the geodesic in $B(\phi_{0}, 2r)\subset M$ given by $c:t\to \exp_{\psi}(t\psi_{*}\eta )$. Along $c$ we have
$$
\frac{d^{2}}{dt^{2}}\,[\exp_{\psi}(t\psi_{*}\eta )]^{i}+(\psi_{*}\eta)^k(\psi_{*}\eta)^j\,\Gamma^i_{jk}(\exp_{\psi}(t\psi_{*}\eta ))=0\;,
$$
where $\Gamma^i_{jk}(\exp_{\psi}(t\psi_{*}\eta ))$ are the Christoffell symbols of the Riemannian connection of $(M,g)$, evaluated at the point $\exp_{\psi}(t\psi_{*}\eta )\in B(\phi_{0}, 2r)$. This yields the expansion
$$
[\exp_{\psi}(\psi_{*}\eta )]^{i}=\psi^i+(\psi_{*}\eta)^i-\frac{1}{2}\,(\psi_{*}\eta)^k(\psi_{*}\eta)^j\,\Gamma^i_{jk}(\psi)+\mathcal{O}(|\psi_{*}\eta|^3)\;.
$$
Substituting into 
\begin{eqnarray}
&&S_{\psi}[\eta_k;\,a^{-1}\,g]\doteq S[\exp_{\psi}(\psi_{*}\,\eta_k);\,a^{-1}\,g]=\\
\nonumber\\
&&a^{-1}\,\int_{\Sigma }\gamma^{\mu\nu}\partial_{\mu}(\exp_{\psi}(\psi_{*}\eta ))^{i}\partial_{\nu}(\exp_{\psi}(\psi_{*}\eta ))^{j}g_{ij}(\exp_{\psi}(\psi_{*}\eta ))d\mu_{\gamma}\nonumber\;,
\end{eqnarray}
and using the expansion
$$
g_{ij}(\exp_{\psi}(\psi_{*}\eta ))=g_{ij}(\psi)+\frac{1}{2}\,(\psi_{*}\eta)^k(\psi_{*}\eta)^l\,[g_{si}(\psi)\nabla _k\Gamma^s_{jl}(\psi)
+g_{sj}(\psi)\nabla_k\Gamma^s_{il}(\psi) ]+\mathcal{O}(|\psi_{*}\eta|^3)\;,
$$
and the definition of the Riemann tensor of $(M,g)$, $\mathcal{R}^{i}_{ksl}=\nabla_k\Gamma^i_{sl}- \nabla_s\Gamma^i_{kl}$, (where 
$\Gamma^i_{sl}$ are interpreted as vector valued endomorphisms), we eventually get 

\begin{eqnarray}
&&S_{\psi}[\eta_k;\,a^{-1}\,g]=a^{-1}\,\int_{\Sigma}\Big[g_{ij}(\psi)\,\gamma^{\mu\nu}\Big(\partial_\mu\psi^i\partial_\nu\psi^j+
2\partial_\mu\psi^i\nabla_\nu(\psi_{*}\eta)^j+\\
\nonumber\\
&&\nabla_\mu(\psi_{*}\eta)^i\nabla_\nu(\psi_{*}\eta)^j\Big)+
R_{ijkl}(\psi)\partial_\mu\psi^i\partial_\nu\psi^l(\psi_{*}\eta)^i(\psi_{*}\eta)^k\Big]d\mu_\gamma +O(|\psi_{*}\eta|^3)\nonumber\;,
\end{eqnarray}
where $\nabla_\mu(\psi_{*}\eta)^i\doteq\partial_\mu(\psi_{*}\eta)^i+(\psi_{*}\eta)^j\partial_\mu\psi^k\Gamma^i_{jk}(\psi)$ is the pullback of the Levi-Civita connection of $M$ to $\psi^*TM$.\\
\\
Since we have approximated the action $S_{\psi}[\eta;\,a^{-1}\,g]$ to second order in $\psi_{*}\eta$, there will be no vertexes with 3 or more legs in the $\eta$-field theory described by (\ref{effaction}).  This implies in particular that no vacuum 1PI--graphs are possible. Thus (\ref{effaction}) reduces to 
\begin{eqnarray}
&&\Gamma_{(0)}(\psi) = a^{-1}\,\int_{\Sigma}g_{ij}(\psi)\,\gamma^{\mu\nu}\partial_\mu\psi^i\partial_\nu\psi^jd\mu_\gamma-\label{effzero}\\
\nonumber\\
&&\ln\Big(\int\exp\Big\{-\frac{1}{2\,a}\int_\Sigma\Big(g_{ij}(\psi)\,\gamma^{\mu\nu}\nabla_\mu(\psi_{*}\eta)^i\nabla_\nu(\psi_{*}\eta)^j+\\
\nonumber\\
&&\gamma^{\mu\nu}
R_{ijkl}(\psi)\partial_\mu\psi^i\partial_\nu\psi^l(\psi_{*}\eta)^i(\psi_{*}\eta)^k     \Big)d\mu_\gamma\Big\}D^{\psi}_g[\eta]\Big)\nonumber\;.
\end{eqnarray}
The $D^{\psi}_g[\eta]$--integration in $\Gamma_{(0)}(\psi)$ gives rise to a functional determinant which, at face value, is divergent. To make sense of it,  one has expand it in powers of $a$, extract the divergent part to each order and eliminate  it by an opportune redefinition of the metric. For illustrative purposes related to Ricci flow theory, we shall do this to first order in $a$, again adapting to our geometrical setting the presentation in \cite{Gaw}.\\
\\
Let $\{e_a\}$ denote a local orthonormal frame in $\psi^{*}TM|_{B(\phi_0,2r)}$, obtained by pulling back an orthonormal frame $\{E_a\}$ defined over $B(\phi_0,2r)$. For notational ease, we shall write $\eta^a$ for the components of $\psi_{*}\eta$ with respect to this $\{e_a\}$. The 
functional integral in (\ref{effzero})  then becomes 
\begin{eqnarray}
\label{gaussin}
\int\exp\Big\{-\frac{1}{2 a}\int_\Sigma(\eta^a\triangle\eta_a+2 (A^\mu)^a_b\eta^b\partial_\mu\eta_a +\\
\nonumber\\
 (A^\mu)^a_b(A_\mu)^c_a\eta^b\eta_c+R_{ajbl}\partial^\mu\psi^j\partial_\mu\psi^l\eta^a\eta^b)d\mu_\gamma\Big\}D^{\psi}_g[\eta]\;, \nonumber
\end{eqnarray}
where we have integrated by parts in the first term, and where $(A_\mu)^a_b$ are the $\{e_a\}$--components of the pullback connection on $\psi^{*}TM|_{B(\phi_0,2r)}$. The Gaussian measure 
$$
D^{\psi}_g[\eta]\,\exp\,-\frac{1}{2 a}\int_\Sigma\eta^a\triangle\eta_a\,d\mu_{\gamma}\,
$$ 
yields a massless field propagator $\langle\eta^a(x)\,\eta^b(y)\rangle$, whereas the remaining terms are treated as interactions. 
The massless field propagator is infrared divergent and needs to be regularized by introducing a small mass term. Geometrically such  a mass term is provided by the natural cut--off distance associated with the fact that we are working with fields $\eta$ 
taking values in $\psi^{*}TM|_{B(\phi_0,2r)}$. Thus defining  $\Lambda '\doteq (2r)^{-1}$, we set 
\begin{equation}
\langle\eta^a(x)\eta^b(y)\rangle = 2\,a\,\frac{\delta^{ab}}{\pi}\int d^2k\frac{e^{ik\cdot(x-y)}}{k^2+\Lambda'^2}
\end{equation}
and let $\Lambda'\to 0$, (\emph{i.e.} $r\to \infty$), at the end. Expanding (\ref{gaussin}) in Feynman graphs using the above propagator on internal lines and three types of  2-legs vertices: \emph{(i)} $A^\mu\partial_\mu$,\,\, \emph{(ii)} $A^\mu A_\mu$ and \emph{(iii)} ${Rm}\;\partial^\mu\psi\partial_\mu\psi$, we find to 1 loop (\emph{i.e.}, to first order in $a$), three divergent graphs $\Upsilon $.
The first, $\Upsilon_{(i)} $ is a loop with two distinct type \emph{(i)} vertices $x$ and $y$. Its Feynman amplitude is given by
\begin{equation} 
F(\Upsilon_{(i)}) = \frac{2}{(a)^2}\int_\Sigma d\mu_\gamma(x)\int_\Sigma d\mu_\gamma(y)(A^\mu)^a_b(x)(A^\nu)^c_d(y)\langle\eta^b(x)\partial_\mu\eta_a(y)\rangle\langle\eta^d(y)\partial_\nu\eta_c(x)\rangle
\end{equation}
The second graph $\Upsilon_{(ii)}$ is a loop with a type \emph{(ii)} vertex $x$ with amplitude
\begin{equation}
F(\Upsilon_{(ii)})= \frac{1}{2a}\int_\Sigma d\mu_\gamma(x) (A^\mu)^a_b(A_\mu)^c_a\langle\eta^b(x)\eta_c(x)\rangle
\end{equation}
Even though each one of these two amplitudes is divergent, their sum is finite, corresponding to the well-known result that there are no 1-loop divergences in 2-dimensional gauge theories. So we are left only with last graph, $\Upsilon_{(iii)}$, associated with a loop with a type \emph{(iii)} vertex $x$:
\begin{equation}
F(\Upsilon_{(iii)})=\frac{1}{2a}\int_\Sigma d\mu_\gamma(x)R_{ajbl}(\psi(x))\partial^\mu\psi^j(x)\partial_\mu\psi^l(x)\langle\eta^a(x)\eta^b(x)\rangle
\end{equation}
We regularize this integral by putting a cutoff $\Lambda$ in the space of momentums. Again such a cut--off has a geometrical origin in the fact that we wish to integrate over $\eta$--fields which confine the corresponding $\phi$--fields in the geodesic ball $B(\phi_0,2r)$, \emph{i.e.} we require that $|k|\leq \Lambda$, with $\Lambda ^{-1}<2r$. Thus 
\begin{eqnarray}
F(\Upsilon_{(iii)})&= \frac{1}{2a}\int_\Sigma d\mu_\gamma R_{ajbl}(\psi)\partial^\mu\psi^j\partial_\mu\psi^l\;2\,a\,\frac{\delta^{ab}}{\pi}\int_{|k|\leq\Lambda}d^2k\frac{1}{k^2+\Lambda'^2}=\nonumber\\
\nonumber\\
&=\ln\left(\frac{\Lambda^2}{\Lambda'^2}\right)\int_\Sigma R_{ij}(\psi)\partial^\mu\psi^i\partial_\mu\psi^j d\mu_\gamma
\end{eqnarray}
Then the 1PI effective action (\ref{effzero}) becomes
\begin{eqnarray}
\Gamma_{(0)}(\psi) = &\int_\Sigma g_{ij}(\psi)\partial^\mu\psi^i\partial_\mu\psi^jd\mu_\gamma +\nonumber\\
&+\,a\,\left(\ln\left(\frac{\Lambda^2}{\Lambda'^2}\right)\int_\Sigma R_{ij}(\psi)\partial^\mu\psi^i\partial_\mu\psi^j d\mu_\gamma +
{finite\quad part}\right)+ \mathcal{O}(\alpha'^2)\;,
\label{finpart}
\end{eqnarray}
where \emph{finite part} indicates terms that are not singular in the limit $\Lambda/\Lambda'\to\infty$. The standard procedure now consists in regarding the metric $g$ in the first term of (\ref{finpart}) as formally infinite and extracting from it a divergent part so to cancel the 1-loop singularity:
\begin{equation}
g_{ij}(\psi) = g_{ij}(\Lambda/\Lambda') -2\,a\,\ln(\Lambda/\Lambda')\,R_{ij}(\psi) + O(a^2)\;.
\label{renmet}
\end{equation}
The metric $g(\psi)$ in the left hand side is the bare metric and $g(\Lambda/\Lambda')$ is the renormalized metric. $R_{ij}(\psi)$ is the Ricci tensor of the bare metric, but \emph{we can as well substitute it with that of the renormalized metric}, $R_{ij}(\psi)\Leftrightarrow R_{ij}[g(\Lambda/\Lambda')]$, since the two metrics are equal to order 0 in $\alpha'$. Substituting (\ref{renmet}) into (\ref{finpart}) we finally get
\begin{equation}
\Gamma_{(0)}(\psi)= \int_\Sigma g_{ij}(\Lambda/\Lambda')\,\partial^\mu\psi^i\partial_\mu\psi^j d\mu_\gamma + a\,({finite\,\,\, part}) + \mathcal{O}(a^2)
\end{equation}
Notice that this procedure does not depend explicitly on the point $\phi_0$ in the geodesic neighborhood of which, $B(\phi_0,2r)$,  we are working, \emph{i.e.} the splitting 
(\ref{renmet}) of the bare metric can be extended smoothly to all $M$. Thus, one can extend the above result to the full nonlinear sigma model (that is to background fields $\psi$ taking values in a geodesic neighborhood  $B(\phi_0,2r)$ of any point $\phi_0$).\\
\\
The renormalizability of the theory depends on the behavior of $g(\Lambda/\Lambda')$ when $\Lambda/\Lambda'\to\infty$; this behavior is described by the beta function (\ref{betaflow}), that we can easily compute from (\ref{renmet}). Indeed, by defining $\tau\doteq \ln(\Lambda/\Lambda')$, we immediately get
\begin{equation}
0=\frac{\partial}{\partial\tau}g_{ij} = \frac{\partial}{\partial\tau}\,g_{ij}(\tau)-2\alpha'R_{ij}(g(\tau)) +\mathcal{O}(a^2)\;.
\end{equation}
Introducing the parameter $t\doteq-a\,\tau$, so that $\partial_tg$ has the same dimension of Ric, one can conclude that the RG flow of the nonlinear sigma model at one loop is \cite{DanThesis}
\begin{equation}
\frac{\partial}{\partial t}\,g(t) = -2{Ric}(g(t))+\mathcal{O}(a^2)\;.
\end{equation}
At this point, it is important to recall that a more detailed analysis at two loops would have produced 
\begin{equation}
\frac{\partial }{\partial t}\,g_{ik}(t)=-2\,R_{ik}(t) \,-\,a\,(R_{ilmn}R_{k}^{lmn})+\,\mathcal{O}(a^{2})\;.
\label{2loops}
\end{equation}
Both these RG flow expressions, in the weak coupling limit $a\to 0$, become R. Hamilton's Ricci flow
(R. Hamilton, '82) \cite{11}
\begin{equation}
\frac{\partial }{\partial t }g_{ab}(t )=-2{R}_{ab}(t )\,,\quad g_{ab}(\eta =0)=g_{ab}\;. 
\label{rf}
\end{equation}
Geometrically this is the weakly--parabolic geometric evolution equation obtained by deforming a Riemannian metric $g_{ab}$, on a smooth $n$--manifold $M$, in the direction of its Ricci tensor $\mathcal{R}_{ab}$ \cite{6,11,12,Huisken}. It must be stressed that the evolution in (\ref{rf}) is weakly parabolic only in the infrared regime for the RG flow, corresponding to $t\to\infty$, whereas the limit $\Lambda/\Lambda'\to\infty$ corresponds to the backward parabolic regime $t\to-\infty$. In particular, the nonlinear sigma model is renormalizable (\emph{i.e.} exists as a continuum theory) iff, starting from the bare metric $g$, we can Ricci-flow backwards in time up to $t=-\infty$ without encountering singularities. In this connection it is also important to note that if, along its  evolution, the Ricci--flow metric develops, somewhere, a region of
large curvature, then the correspondence between the $\mathcal{RG}$--flow and the Ricci flow breaks down. In such a case
one needs to consider at least the $a\,(R_{ilmn}R_{k}^{lmn})$ term in (\ref{2loops}), and the large--distance behavior ($t\to+\infty$) may strongly depend also on topological terms to be added to $S[\phi;a^{-1}g]$. Conversely, the development of singularities as $t$ decreases imply that we cannot remove the UV--cutoff $\Lambda$. The action $S[\phi;a^{-1}g]$ does not define a local field theory, and the best one can hope for is an
effective description valid at some scale $t_{0}$.\\

\section{The geometry of Ricci flow}

Ricci flow has been the point of departure and the motivating example for important developments in geometric analysis, most spectacularly for G. Perelman's proof \cite{18, 19, 20} of the  Thurston geometrization program for three-manifolds \cite{thurston1,thurston2} and of the attendant Poincar$\acute{e}$ conjecture. Thus, it may appear really amazing that Ricci flow comes out so naturally from the renormalization group analysis of non--linear sigma models. Since QFT indicates that (\ref{rf}) is just the weak coupling approximation to the full RG flow, there are also potentially useful implications in such a natural QFT ambientation: the renormalization group approach may indeed points towards possible generalizations that may bypass known shortcomings of the Ricci flow. Is there a geometrical fact working behind the scene which can be held responsible of this ubiquitous role of the Ricci flow both in geometry and RG analysis? The answer is in the affirmative and it is intimately connected to the fact that the beta function $\beta(\alpha(t))$ for the RG flow  can be seen as a vector field in the relevant coupling space $\mathcal{C}$ of the given QFT. As we have seen, in the case of non linear sigma model, a basic component of this coupling space is provided by the space of Riemannian structures $\frac{\mathcal{M}et(M)}{\mathcal{D}iff(M)}$, and the answer to the above question lies in the fact that this space comes naturally endowed with a distinguished vector field which is proportional to the Ricci tensor. 
To explain what this statement means, let us assume that $M$ is a $C^{\infty }$ compact  manifold without boundary, and let us denote by $C^{\infty }(M, {R})$ and 
${C}^{\infty }(M,\otimes ^{p}\, T^{*}M\otimes ^{q}TM )$ the space of smooth functions and of smooth $(p,q)$--tensor fields over 
$M $, respectively.  
Recall that we have denoted by $\mathcal{D}iff(M )$ the group of smooth diffeomorphisms of $M $, and by $\mathcal{M}et(M )$ the space of all smooth Riemannian metrics over $M$. The tangent space\,, $\mathcal{T}_{(M ,g)}\mathcal{M}et(M )$, to $\mathcal{M}et(M )$ at $(M,g )$ can be naturally identified with the space of symmetric bilinear forms ${C}^{\infty }(M,\otimes ^{2}_{S}\, T^{*}M)$ over $M $, endowed with the pre--Hilbertian  $L^{2}$ inner product  $(U,V)_{L^{2}(M )}\doteq \int_{M }g^{il}\,g^{km}\,U_{ik}\,V_{lm}d\mu _{g}$ for  $U,\,\,V\,\in\,{C}^{\infty }(M,\otimes ^{2}_{S}\, T^{*}M)$. 
Let ${L}^{2}(M,\otimes ^{2}\, T^{*}M)$  be the corresponding $L^{2}$ completion of ${C}^{\infty }(M,\otimes ^{2}_{S}\, T^{*}M)$. An important geometric property of $\mathcal{M}et(M )$ is that the tangent space $\mathcal{T}_{(M ,g)}\mathcal{O}_{g}$ to the $\mathcal{D}iff(M )$--orbit of a given metric $g\in \mathcal{M}et(M )$ is the image of the injective operator $\delta_{g} ^{*} \,:C^{\infty }(M ,T^{*}\,M )\rightarrow  C^{\infty }(M ,\otimes ^{2}T^{*}\,M)$ defined by $\delta_{g}^{*}\,(w_{a}\,dx^{a})\doteq \frac{1}{2}\,\mathcal{L}_{w^{\#}}\,g$ where $\mathcal{L}_{w^{\#}}$ is the  Lie derivative along the vector field $(w^{\#})^{i}\doteq g^{ik}w_{k}$. Standard elliptic theory then implies that the $L^{2}$--orthogonal subspace to $Im\;\delta_{g} ^{*}$ in $\mathcal{T}_{(M,g)}\mathcal{M}et(M )$ is spanned by the ($\infty $--dim) kernel of the $L^{2}$ adjoint $\delta_{g}$ of  $\delta_{g} ^{*}$, defined by $\delta_{g} \,(h_{ab}\,dx^{a}\otimes dx^{b})\doteq -\,g^{ij}\,\nabla _{i}h_{jk}\,dx^{k}$.
It follows that with respect to the inner product $(\circ ,\circ )_{L^{2}(M )}$, the tangent space $\mathcal{T}_{(M ,g)}\mathcal{M}et(M )$ splits as \cite{ebin} $\mathcal{T}_{(M ,g)}\mathcal{M}et(M )\cong  Ker\;\delta_{g} \,\oplus Im\;\delta_{g} ^{*}$. This $L^{2}$ splitting of $\mathcal{T}_{(M ,g)}\mathcal{M}et(M )$ implies that,
unless $\mathcal{R}ic(g)\equiv C\,g+\mathcal{L}_{w^{\#}}\,g$ for some vector field $w^{\#}$ and some constant $C$, the Ricci tensor $\mathcal{R}ic(g)$ of a metric $g\in \mathcal{M}et(M )$ can be thought of as a non--trivial $\mathcal{D}iff(M )$--equivariant section of the tangent bundle $\mathcal{T}\,\mathcal{M}et(M )$, \emph{i.e.}, $\{\mathcal{R}ic(g)\}\cap Ker\;\delta_{g}\not=\emptyset $. Thus, the Ricci flow associated with a Riemannian three-manifold $(M,g)$ can be thought of as the $\mathcal{D}iff(M )$--equivariant dynamical system on $\mathcal{M}et(M )$ generated by the weakly-parabolic diffusion--reaction PDE \cite{11}
\begin{eqnarray} 
&&\frac{\partial }{\partial t }g_{ab}(t )=-2\mathcal{R}_{ab}(t)\,,\label{mflow}\\
\\ 
&&g_{ab}(t =0)=g_{ab}\,\quad 0\leq t <T_{0}\nonumber\;,
\end{eqnarray}
where $\mathcal{R}_{ab}(t )$ is the Ricci tensor of the metric $g_{ik}(t )$.
It follows from the above characterization that the geometrical and analytical properties featuring in the Ricci flow are the study of  non--linear parabolic systems of PDEs and the structure theory for Riemannian manifolds. In this connection,  let us recall that the flow $(M ,g) \mapsto (M ,g(t ))$, defined by (\ref{mflow}), always exists in a maximal interval $0\leq t < T_{0}$,  
for some $T_{0}\leq \infty $. If such a $T_{0}$ is finite then 
$\lim_{t \nearrow T_{0}}\, [\sup_{x\in M }\,|Rm(x,t )|]=\infty $, \cite{11,12} where $Rm(t )$ is the Riemann tensor of $(\Sigma ,g(t ))$. 
Conversely, if $\lim_{t \nearrow T_{0}}\, [\sup_{x\in M }\,|Rm(x,t )|]<\infty $, then the solution can be uniquely extended past time $T_0$. Among  solutions of the Ricci flow an important role, (typically in singularity analysis as well as in related physical applications of the theory), is played by  generalized fixed points associated  with the action on $\mathcal{M}et(M)$ of $\mathcal{D}iff(M )\times {R}_{+}$, where ${R}_{+}$ acts by scalings. These solutions are described by the \emph{Ricci solitons} $-2{\mathcal{R}}_{ab}(t )=\mathcal{L}_{\vec{v}(t )}\,{g}_{ab}+\, \varepsilon \,{g}_{ab}$, where  $\mathcal{L}_{\vec{v}(t )}$ denotes the Lie derivative along the $t $--dependent (complete) vector field ${\vec{v}(t )}$ generating $t \mapsto \varphi (t )$\,$\in\,\mathcal{D}iff(M )\times {R}_{+}$, and where, up to rescaling, we may assume that $\varepsilon =\,-1,\; 0,\; 1$, (respectively yielding for the \emph{shrinking, steady, and expanding solitons}). 

\section{Remarks on singularities of the Ricci flow}

The understanding of how the solutions to the Ricci flow look as they approach a singular regime is a key step in exploiting the Ricci flow in the proof of the geometrization conjecture. Moreover, according to what we have seen in the analysis of RG for non linear sigma models, it is evident that also in such a setting singularity formation plays a basic role. In this latter case the relevant solutions of the Ricci flow are the \emph{ancient solutions}, the ones which exists on a maximal time interval $-\infty<t<T_0$, where $T_0<\infty$. These correspond to renormalizable sigma-models.
As suggested by R. Hamilton \cite{12}, a natural classification of singularities can be based on how long the solution to the Ricci flow exists and how such a solution scales asymptotically. Let us recall that if a solution $t\mapsto g_{ab}(t)$, $0\leq t< T_{0}$, to the Ricci flow develops a singularity at the maximal time $T_{0}$, then such a singularity is said to be a Type--$I$ singularity (rapidly forming) if $\sup_{t\in [0,T_{0})}(T_{0}-t)\,\mathcal{K}_{max}(t)<+\infty $, whereas it is said to be a Type--$II_{a}$ singularity (slowly forming) if $\sup_{t\in [0,T_{0})}(T_{0}-t)\,\mathcal{K}_{max}(t)=+\infty $, where $\mathcal{K}_{max}(t)$ $\doteq $ $\sup_{x\in M }\{|\mathcal{R}m\,(x,t)|\}$. We have a similar classification for infinite time singularities:  
a singularity is said to be a Type--$III$ singularity (rapidly forming) if $\sup_{t\in [0,\infty )}\,t\,\mathcal{K}_{max}(t)<+\infty $, whereas it is said to be a Type--$II_{b}$ singularity (slowly forming) if $\sup_{t\in [0,\infty )}\,t\,\mathcal{K}_{max}(t)=+\infty $. In particular, for ancient solutions we say that $(M,g(t))$ is a   Type--$I$ ancient solution if $\sup_{t\in (-\infty,-1]}\,|t|\,\mathcal{K}_{max}(t)<+\infty $, and is 
a   Type--$II$ ancient solution if $\sup_{t\in (-\infty,-1]}\,|t|\,\mathcal{K}_{max}(t)=+\infty $.\\
In Ricci flow theory there is a standard technique, connected to parabolic rescaling, that is used to study what happens as the Ricci flow approaches a singularity. This is known as \emph{point picking} (\emph{e.g.}, \cite{Ni} p.297):
Assume that $t\rightarrow (M ,g(t))$ is a solution to the Ricci flow defined on a maximal time interval $[0,T)$, where $T\leq \infty $, so that $\sup_{M \times [0,T)}\,|Rm|=\infty $ if $T<\infty $. In order to  understand the singularity which is forming as $t\rightarrow T$, one considers a sequence of points $x_{i}\in M$ and times $t_{i}\nearrow T$, and out of the given solution $t\rightarrow (M ,g(t))$ one constructs a sequence of solutions $t\rightarrow (M ,g_{i}(t))$ defined by $g_{i}(t)\doteq K_{i}\,g(t_{i}+K_{i}^{-1}\,t)$, where  $K_{i}\doteq |Rm(x_{i},t_{i})|$. The interesting sequences for singularity formation are those for which $\lim_{i\rightarrow \infty }\,K_{i}=\infty $, but note that on any fixed compact time interval the curvature of the metrics $g_{i}(t)$ are bounded. This rescaling technique naturally opens the way to the application of Gromov--Hausdorff techniques in Ricci flow theory, and indeed R. Hamilton was able to prove a compactness theorem, (uniform convergence in $C^{\infty }$ on compact sets), for solution to the Ricci flow.
The hypotheses under which Hamilton's result holds  require in an essential way a control (uniform lower boundedness) of the injectivity radiuses of 
$t\rightarrow (M ,g_{i}(t))$ at framed marked points $\{O_{i},F_{i}\}$, where $F_{i}$ denotes an orthonormal frame (with respect to  $(M ,g_{i}(t=0))$) at $O_{i}\in M$. Such a control is not a natural consequence of the parabolic rescaling technique described above and important developments in Ricci flow theory have been strictly connected to devising ways of proving that such a control on the injectivity radius is naturally associated with solutions of the Ricci flow.\\
Removing the requirement on the injectivity radius is equivalent  to considering  collapsing sequences for solutions of the Ricci flow, and it is indeed possible \cite{David} to extend Hamilton's compactness theorem to collapsing sequences of metrics via a study of how the size of metric balls evolve under the  Ricci flow. This is very much in the spirit of RG group, since the delicate part of the proof is to understand the local geometry of the resulting limit space in a (pointed) Gromov--Hausdorff topology by describing the local geometry of collapsed limits of a sequence of Riemannian manifolds.  Recently, such results have been extended by John Lott and David Glickenstein
\cite{lottgroup,David2}, who using the technique of Riemannian groupoids have been able to piece together the local limit flows and show how one produces solitons in  the limit. This has provided an understanding of the nature of the Type--III singularities that occur in the case of compact homogeneous geometries with a complete classification of the relevant 3--dimensional case. It is important to stress that this analysis builds up on
a way of discussing collapsing limit of  Ricci flow on a space which basically has the same dimension as the original manifold. Collapsing takes the form of a symmetry (the collapsing symmetry) under which the limit Ricci flow is equivariant. A phenomenon which appears similar to the mechanism generating a quantum field out of of a broken symmetry. Without going into detail, under the action of this collapsing symmetry, the limit Ricci solution takes the structure of a Riemannian groupoid. This is a (categorial) notion familiar in foliation theory and in the study of the $C^{*}$--algebra of a foliation. Roughly speaking, the notion of Riemannian groupoid unifies in a unique definition the notion of manifold, orbifold, and quotient manifold. As such it appears of great potential relevance in the study of the UV regime in the RG flow for non linear sigma models.\\
\\
It is clear that the study of the nature of singularity formation is one of the main topic of interest in Ricci flow theory since it provides an understanding of the structure of solutions in high curvature regions. Relevant topological information about the long--time behavior of Ricci flow solutions has been provided by Perelman in his celebrated papers, however it is more than fair to admit that explicit and precise quantitative information is still missing. According to the analogies drawn here it is possible that the mathematical imagination of QFT may suggest new strategies where the geometry, transcending and transgressing formal boundaries, is perhaps more comfortable.

\section*{Acknowledgements}

M.C. wish to express his gratitude to the organizers of the Symposium for the kind ospitality and the friendly atmosphere in Torun.

\end{document}